\documentclass{ws-ijmpcs}

\usepackage{latexsym,bm,amsmath,amssymb,amsfonts, dsfont}
\usepackage{epsfig,graphics,graphicx,mathrsfs}
\usepackage{slashed}
\usepackage{cite}

  \newcommand{\abar}{\bar{\alpha}_s}
  \newcommand{\lan}{\left\langle}
  \newcommand{\ran}{\right\rangle}
  \newcommand{\mcal}{\mathcal}
  \newcommand{\beq}{\begin{eqnarray}}
  \newcommand{\eeq}{\end{eqnarray}}

\begin{document}

\markboth{Emil Avsar}
{On the understanding and use of ``unintegrated'' parton distributions in small-x QCD}

%
\catchline{}{}{}{}{}
%

\title{ON THE UNDERSTANDING AND USE OF ``UNINTEGRATED'' PARTON DISTRIBUTIONS IN SMALL-X
QCD}

\author{EMIL AVSAR}

\address{Department of Physics, Penn State University,\\
University Park, PA 16802,
USA}



\maketitle


\begin{abstract}
We review and discuss the use of TMD, or "unintegrated", gluon distributions in the
domain of small-$x$ physics. The 
definitions employed, and the hazards of the naive applications of the TMD factorization 
and the associated gluon distributions are discussed. 
  
\end{abstract}

\ccode{PACS numbers:  12.38.Aw, 12.38.Bx, 24.85.+p}

\section{Introduction}	

The concept of transverse-momentum-dependent (TMD), or $k_\perp$-dependent, or 
``unintegrated'', parton distributions is frequently encountered in the QCD literature.  This is so  
particularly in the small-$x$ domain where they play a 
prominent role.  These distributions are to be contrasted with the integrated parton distributions 
(parton distribution functions, pdfs)  which play a fundamental role in global QCD fits.  

For the $k_\perp$-dependent distributions there is no unique definition, and as a consequence 
it is for the phenomenology of these quantities very important to relate the different definitions.  
We shall therefore here analyze some of the most commonly found definitions in the literature. 

Before going any further, we should mention that there exist two different and distinct
sets of users of these distributions, each with corresponding domains of application:

\begin{itemlist}
\item One area we might characterize as ``hard scattering factorization'',
  where the primary object of attention is a hard scattering
  factorization property.  An example is the Drell-Yan process at
  low transverse momentum. 
\item The other area is that of small $x$,  DIS with
  $x\ll1$ being a typical example.  Here the
  emphasis starts with the BFKL  formalism (or a generalization of it) for scattering
  processes in the Regge region.  
  \end{itemlist}

We are here primarily interested in the small-$x$ domain. The theory of the ``hard scattering'' domain is 
in great detail explained in \cite{qcdbook} 
(see also \cite{Collins:2003fm} for a short review),  and a compact presentation is also given 
in the talk by John Collins \cite{Johntalk}. We also note that the terminology ``TMD'' distribution is usually used 
in the ``hard scattering factorization'' while traditionally in the small-$x$ domain one speaks of ``$k_\perp$-dependent'', 
or more commonly of ``unintegrated'' parton distributions.  

\section{The parton model as background}

In the parton model, the concept of 
a parton distribution quantifies the intuitive expectation of a number density of partons of given flavor
in the target hadron.  A mathematically exact definition can be 
given in the light-front quantization as
\beq
\label{eq:pdf.lf.def}
  f_{j/h}(x,k_\perp) = \sum_\alpha 
         \frac{1}{ 2x (2\pi)^3 }
         \frac{\langle P,h |  a_{k,\alpha,j}^{\dag} a_{k,\alpha,j} |P,h\rangle}
              {\langle P,h | P,h\rangle}.
\eeq
Here $j$ and $h$ label parton and hadron flavor, $\alpha$ is a parton
helicity index, $|P,h\rangle$ is the target state of momentum\footnote{Strictly speaking 
the momentum states $|P\rangle$ do not belong to the ordinary Hilbert space of states since they 
are not normalizable. One can remedy this by replacing them with wave packets, and then taking appropriate 
limits as these packets approach the momentum states, but we need not worry about such level of rigor here.} $P$, and
$a$ and $a^\dagger$ are parton light-front annihilation and creation operators.  We use light-front coordinates,
the hadron target has zero transverse momentum, and $x=k^+/P^+$.  Integrating Eq. ~\eqref{eq:pdf.lf.def} over
\emph{all} $k_\perp$ one obtains the integrated distribution
\beq
 f_{j/h}(x)= \int_{\mathrm{all}\,\, k_\perp} \!\!\!\! d^2k_\perp \, f_{j/h}(x,k_\perp).
\label{intvsunintpm}
\eeq
The exactness of the 
relation, however, depends on the specific assumptions of the parton model which no longer hold
in full QCD. 

A superficial glance over the relevant literature reveals that the number density interpretation 
of the TMD, $k_\perp$-dependent, and unintegrated parton distributions is taken rather literally also 
in QCD. The actual technical definitions in the different cases, however,  
may or may not conform to the idea of a number density. We also note that the parton model relation 
between the integrated and the TMD distributions in \eqref{intvsunintpm} is taken to be true in QCD as well 
(with the difference that the $k_\perp$ integral is performed only up to the hard scale $Q$). 
The question then is what exactly the justifications are for these assumptions. 

\section{The gluon distribution at small-$x$}

\subsection{BFKL and the dipole picture}

The prototype of the QCD applications in the small-$x$ domain is the BFKL formalism \cite{Fadin:1975cb, Kuraev:1977fs, Balitsky:1978ic}. Here the $\gamma^*\gamma^*$ scattering amplitude (for $A+B\to A'+B'$) is in the Regge limit ($s/t \gg 1$ with $t=-q_\perp^2$) written as
\beq
\frac{\mathrm{Im}A(s,t)}{s} = \int \frac{d^2 k_\perp}{k_\perp^2} \frac{d^2 k_\perp'}{(k_\perp'-q_\perp)^2}
I_A(k_\perp,q_\perp) I_B(k_\perp',q_\perp) F(s, k_\perp, k_\perp', q_\perp),
\label{bfklfact}
\eeq
where $A$ and $B$ denote the two virtual photons. 
The dimensionless objects $I_A$ and $I_B$ are called the impact factors, while  $F$ is the object satisfying the BFKL equation
 (Eq.~\eqref{bfkleq} below), and is commonly referred to as the ``BFKL Greens' function''.  If by $\phi_A$ and $\phi_B$ one denotes the couplings
of gluons to the external photons $A$ and $B$ (via quark boxes), then to lowest order in the singlet channel in Feynman gauge (two gluon exchange) the amplitude can be written in shorthand notation as 
\beq
\phi_A^{\mu\nu}\phi_B^{\alpha\beta}g_{\mu\mu'}g_{\nu\nu'}g_{\alpha\alpha'}g_{\beta\beta'}
\langle 0|T \, A^{\mu'}A^{\nu'}A^{\alpha'}A^{\beta'} |0\rangle  \nonumber  \\
\approx_{eik} I_A\, I_B \, \langle 0|T \,( p_A\cdot A) \, (p_A\cdot A) \,( p_B \cdot A) \,( p_B\cdot A) |0\rangle. 
\label{loworder}
\eeq
Here $\approx_{eik}$ indicates the eikonal approximation in which the numerator of the Feynman gauge 
propagator $g^{\mu\nu}$ is replaced by $p_B^\mu p_A^\nu/p_A\cdot p_B$ ($p_A^\mu p_B^\nu/p_A\cdot p_B$)
for gluons coupling to $A$ ($B$).  Therefore to lowest order $F$ is given by the vacuum expectation value of four off-shell 
gluons. To all orders, the gluon fields have to be summed into eikonal gauge links, \emph{i.e} Wilson lines, which 
make $F$ gauge invariant.  

In the space-time picture by Balitsky \cite{Balitsky:1995ub}, the $q\bar{q}$ pair emerging from the electromagnetic currents $J$
from one of the particles (say $A$) travels along a straight line trajectory and scatters off the gluon field created 
by $B$ which has a delta function shape, \emph{i.e} a shockwave, in the high energy approximation.  
The propagation of the $q\bar{q}$ in the external gluon field created by $B$ is represented by the Wilson 
lines
\beq
U_A(x_\perp) = P \exp \left( -i g \int_{-\infty}^\infty d \lambda \, n_A \cdot A^a(x_\perp + \lambda n_A )t_F^a
\right ),
\label{Wilsonsmallx}
\eeq
where the color matrix $t_F^a$ is in the fundamental representation, and $n_A$ is a unit vector in the  direction 
of $p_A$.
This picture also leads to the so-called
dipole formalism, as the $q\bar{q}$ pair emerging from the current $J$ is in a color singlet state and thus can be 
seen as a ``color dipole''.   The lowest order result in Eq.~\eqref{loworder} is in the formulation by Balitsky generalized 
to 
\beq
\frac{\mathrm{Im}A(s,t)}{s} =  \int d^2 k_\perp I_A(k_\perp, q_\perp) 
\lan \mathrm{Tr} \{U_A^\dagger \, U_A\}(k_\perp, q_\perp)  \ran 
\label{balitsky3}
\eeq
where
\beq
\lan  \mathrm{Tr} \{U_A \, U_A^\dagger\}(k_\perp, q_\perp)  \ran  &\equiv& 
\int d^2x_\perp e^{-i k_\perp \cdot x_\perp} \lan  \mathrm{Tr} \{U_A^\dagger(x_\perp) 
U_A (0_\perp) \} \ran_{q_\perp} \!. 
\eeq
The exact definition of the bracket $\lan \mathcal{O}  \ran$ is given in \cite{Balitsky:1995ub}.
Here it suffices to say that it involves a convolution of the Wilson lines with the currents $J_B$ and $J_{B'}$.  
Notice that the Wilson lines here correspond to the all order summation of the factors $p_A\cdot A$ in \eqref{loworder},
while the fields $p_B\cdot A$ are kept implicitly in the definition of the bracket. When made explicit they 
should give rise to an additional pair of Wilson lines $U_B$ in the direction of $p_B$. 


In formula \eqref{bfklfact} the main object is the BFKL function $F$.  From \eqref{balitsky3} we see that 
one has
\beq
\lan \mathrm{Tr} \{U_A^\dagger \, U_A\}(k_\perp, q_\perp)  \ran = 
 \int \frac{d^2 k'_\perp}{k_\perp^2(k_\perp'-q_\perp)^2} I_B(k_\perp',q_\perp) F(s, k_\perp, k_\perp', q_\perp).
 \label{balitskybfkl}
\eeq
The expectation value of the Wilson line pair is essentially taken 
to be the ``gluon distribution'' (also called the ``dipole gluon distribution'') in the small-$x$ domain. 
While the formulas here are derived for $\gamma^* \gamma^*$
scattering, a similar identification of the ``gluon distribution'' is made for other processes as well, such 
as DIS off a proton or a nucleus, and also in proton-proton ($pp$), proton-nucleus ($pA$), and 
nucleus-nucleus ($AA$) collisions. For these latter processes we have not been able to find any 
proofs showing that this is indeed fully legitimate.  
The definition of the ``dipole gluon distribution'' is then generally  taken to be
\beq
\mcal{F}_{\mathrm{dip}} = \mcal{C} \int d^2x_\perp k_\perp^2 \,e^{-i k_\perp \cdot x_\perp} \langle P, h | \mathrm{Tr} \{U^\dagger(x_\perp)U (0_\perp)\} 
|P, h\rangle
\label{dipgluedistrb2}
\eeq
where we leave the prefactor $\mcal{C}$ unspecified since there does not seem to be 
a universally accepted value in the literature\footnote{In fact we have find several incompatible values of $\mcal{C}$
used in different studies of the same processes. Moreover, even the form of \eqref{dipgluedistrb2} varies 
from case to case so it is hard to pin down any exact universal definition.}.  
We should also mention that one needs to insert transverse gauge links 
at infinity to make  the operators in \eqref{balitsky3} and \eqref{dipgluedistrb2} exactly gauge invariant.

We notice that the structure of \eqref{dipgluedistrb2} does not really conform to the parton model 
definition \eqref{eq:pdf.lf.def} of a number density.  Therefore, despite its name, this object does not 
correspond to the direct generalization of the parton model definition  \eqref{eq:pdf.lf.def}.  Moreover, 
the Wilson lines in the definition are always take in the fundamental representation while the 
generalization of the parton model result to full QCD naturally leads to Wilson lines in the adjoint 
representation for the gluon distribution \cite{qcdbook}. 
It is, however, still common to assert the relation \eqref{intvsunintpm} to the integrated pdf, by 
writing
\beq
G(x,Q^2) = \int^{Q^2} d^2k_\perp  \mathcal{F}_{\mathrm{dip}}(x, k_\perp).
\label{intvsunintqcd}
\eeq
By $G$ we here denote the gluon pdf.  For an explanation of the meaning of the parameter $x$ in $\mathcal{F}_{\mathrm{dip}}$, 
see below.

A reason for asserting this result
is that the leading twist approximation of the BFKL evolution, which $\mathcal{F}_{\mathrm{dip}}$ satisfies, reproduces 
for $G$, via \eqref{intvsunintqcd}, the DGLAP
evolution in the approximation where $n_f=0$ and where only the $1/z$ term in the splitting function $P_{gg}$ 
is kept.  

Despite this agreement in the simplified limit, 
it is, however, a bad practice to hold on to the relation \eqref{intvsunintqcd}
for several reasons. 
One reason is that the partial agreement of the evolution equations does not generally imply that 
the relation \eqref{intvsunintqcd} 
is a priori consistent with the respective operator definitions. The integrated pdf, $G$, which follows from the collinear
factorization approach has a clear operator definition (see for example \cite{qcdbook}) that cannot simply be 
reduced to 
the integral of \eqref{dipgluedistrb2} (or some variant of it).  Moreover,  it is generally for phenomenology important to include the non-singular parts of the DGLAP splitting function, even at rather small-$x$.
Therefore the simplified limit mentioned above is not in practice very useful. 

There are also conceptual problems. It is known that the use of light-like Wilson lines in the operator 
definitions for the TMD distributions gives rise to rapidity divergences which have to be somehow cut-off. 
A convenient way to cut the 
divergences is to take the Wilson lines to be non-light-like \cite{qcdbook, Balitsky:1995ub}, 
\emph{i.e.} to let them have finite rapidity. 
It is actually this cut-off  which gives rise to 
the evolution in rapidity, just as the cut-off $\mu$ in momenta gives rise to the standard DGLAP evolution 
with respect to $\mu$.  If we denote the cut-off in rapidity by $\zeta$, then $\mathcal{F}_{\mathrm{dip}}$ 
depends on $k_\perp$ and $\zeta$, $\mathcal{F}_{\mathrm{dip}}=\mathcal{F}_{\mathrm{dip}}(k_\perp;\zeta)$. 
It is important to realize that $\zeta$ is conceptually different than the variable $x$ appearing in 
 $f_{j/h}(x,k_\perp)$ in the parton model definition  \eqref{eq:pdf.lf.def}.  In that case $x$ is an 
 ``intrinsic'' or kinematical variable, and is literally the light-cone momentum fraction $x=k^+/P^+$ of the 
 parton which participates in the hard scattering.  On the other hand $\zeta$ is an arbitrary cut-off variable which determines the total rapidity range for the soft gluons associated with  $\mathcal{F}_{\mathrm{dip}}$, and has no counterpart in the parton model.  
 Now, in the small-$x$ literature one always sets 
 $\zeta=x$ (we remind that in DIS $x=Q^2/2q\cdot P$) .  This is the reason behind the notation $\mathcal{F}_{\mathrm{dip}}(x,k_\perp)$. 
 
The variable $x$ in the integrated distribution is not really related to the rapidity cut-off in $\mathcal{F}_{\mathrm{dip}}$. It is in the integrated distribution a kinematical variable which has the meaning of the 
total light-cone momenta $k^+=x\,P^+$ exchanged in the $t$-channel between 
the hard scattering and the target hadron. On the other hand in $\mathcal{F}_{\mathrm{dip}}$
it arises as the rapidity cut-off which can be implemented by taking the Wilson lines $U$ to be non light-like. 
In fact, the corresponding kinematical variable $x$ has in the derivation of \eqref{bfklfact}, \eqref{balitsky3} and \eqref{dipgluedistrb2}
already been set to 0 (that is, the direct analogue of $x$ in \eqref{eq:pdf.lf.def} has already been set to 0 in  \eqref{dipgluedistrb2}).  A more correct notation for $\mathcal{F}_{\mathrm{dip}}$ would therefore be $\mathcal{F}_{\mathrm{dip}}(x=0,k_\perp;\zeta=x)$.
The integrated distribution $G$ does not have any $\zeta$ dependence because it does not contain any rapidity divergences. 
We therefore see that one generally has to be more careful when relating $G$ and $\mathcal{F}_{\mathrm{dip}}$. 


\subsection{The ``Weizsacker-Williams'' distribution}

There is also a different type of gluon distribution found in the small-$x$ literature which is meant 
to literally be the analogue of the parton model definition \eqref{eq:pdf.lf.def}.  This distribution has been 
dubbed the ``Weizsacker-Williams'' (WW) gluon distribution, and is in a sense somewhat more closely related to 
the gluon distribution in the hard scattering factorization approach.  A definition is provided in 
 \cite{Iancu:2002xk} which starts from the relation
\beq
\frac{dN}{d^3k} = \lan a_a^{i\, \dagger}(x^+, k) \, a_a^{i}(x^+, k) \ran = \frac{2k^+}{(2\pi)^3} \lan
A^i_a(x^+, k) \, A^i_a(x^+, -k) \ran
\label{numberdens}
\eeq
where $a$ and $a^\dagger$ are the light-front gluon field annihilation and creation operators. 
Noticing that in light-cone gauge, $A^+ = 0$, one has $F_a^{+i}=ik^+A_a^i$, it is seen that this
definition thus corresponds to the expectation value $\langle F_a^{+i}F_a^{+i} \rangle$. Indeed this is the 
direct generalization of the parton model definition \eqref{eq:pdf.lf.def}.  The result in a generic gauge is 
then taken to be  \cite{Iancu:2002xk}
\beq
\mathcal{F}_{WW}(x,k_\perp;\zeta) =  \frac{2}{(2\pi)^3}
\int dx^-dy^-\!\!\int d^2x_\perp && \!\!\! d^2y_\perp e^{ixP^+ (x^- -y^-) - ik_\perp (x_\perp-y_\perp)} \nonumber \\
&&\lan  F_a^{+i}(x) \tilde{U}_{ab}(x,y)F_b^{+i}(y) \ran_{W_{\zeta \, P^+}}.
\label{WWdistrb}
\end{eqnarray}
Here $\tilde{U}(x,y)$ denotes a Wilson line from $y=(0^+,y^-,y_\perp)$ to $x=(0^+,x^-,x_\perp)$ in the adjoint 
representation.

It should be noted that the definition \eqref{WWdistrb} is classical in the sense that all gluon fields in the expectation value are classical color fields, and the averaging $\langle \mathcal{O} \rangle_{W_{\zeta \, P^+}}$ is performed using the 
classical distribution function $W$ of CGC.  All the effects of quantum fields has been inserted into the evolution 
of $W$ with respect to the cut-off parameter $\zeta$ (indicated by the notation $W_{\zeta P^+}$). The role of 
$\zeta$ is here identical to the cut-off discussed above. In fact, in the CGC, $\zeta P^+$ acts a sharp cut-off so that 
gluons with $k^+ < \zeta P^+$ are removed completely.  Note that we again keep distinct the variables $x$ 
and $\zeta$. 

Notice that the mere requirement of gauge invariance does not fix the paths of the gauge links in 
\eqref{WWdistrb} exactly.  
Thus if the general form of the result is only dictated by gauge invariance, then the paths of the Wilson lines
remain ambiguous.  In the hard scattering factorization, the Wilson lines of the TMD distribution are fixed 
by whatever is needed in order to obtain factorization.  
In the CGC, however, the distribution \eqref{WWdistrb} is not related to any factorization 
theorem. Its existence is simply conjectured from the intuitive expectation of the 
gluon distribution corresponding to a number density. It is furthermore not clear how  \eqref{WWdistrb} arises 
from Feynman graph calculations in the small-$x$ limit since there does not exist any direct derivation of it. 

As the WW distribution is in structure more similar to the TMD distributions in the hard scattering 
factorization approach, it would be interesting to actually find a derivation of it from a factorization 
theorem in the small-$x$ limit. For some related work see also \cite{Dominguez:2011wm}.

\subsection{The CCH approach}

The standard reference to $k_\perp$-factorization in small-$x$ physics is the work by 
Catani, Ciafaloni and Hautmann (CCH) \cite{Catani:1990eg}. One studies here 
the production of heavy $q\bar{q}$ pairs in photoproduction, DIS, and in hadron-hadron
collisions, and the main goal is to formulate a TMD factorization formula at small-$x$ 
which at the same time can in the collinear limit be related to the standard collinear factorization 
formula. 

The factorization formula, based on the properties of the light-cone gauge, is written as (in case of photoproduction)
\beq
\sigma_{\gamma g} = \frac{1}{2M^2}\int \frac{dz}{z} \int d^2k_\perp 
\hat{\sigma}(\rho/z,k_\perp^2/M^2) \, \mathcal{F}(z,k_\perp),
\label{CCHfact}
\eeq
where $\rho = 4M^2/s$, and $M$ is the invariant mass of the heavy quark. This formula is somewhat 
different than the BFKL formula \eqref{bfklfact}, and is more close in spirit to the hard factorization approach since 
$\hat{\sigma}$ here plays the role of a hard scattering coefficient. The object referred to as the 
``unintegrated gluon distribution'' can in the light-cone gauge used in \cite{Catani:1990eg} be written as 
\cite{ourpaper}
\beq
\mathcal{F}(x,k_\perp) = \int \frac{dx^- d^2x_\perp}{(2\pi)^3} \frac{1}{4 \,P^+}\,e^{ix P ^+x^--ik_\perp x_\perp}
\langle P| F_a^{+i}(0^+,x^-,x_\perp) F_a^{+i}(0) |P\rangle.
\label{LCgluondens}
\eeq

This definition exactly corresponds to the parton model one in \eqref{eq:pdf.lf.def} (to $xf_{g/h}$), and is also 
the analogue of the WW distribution in the CGC.  
If taken literally, however, it leads to rapidity divergences (in this case due to the singular denominator
of the light-cone gauge propagator), and consequently it must be modified. The point is, however, that the definition 
is never actually used in \cite{Catani:1990eg}. It is instead stated that $\mathcal{F}(z,k_\perp)$ in \eqref{CCHfact} is ``defined'' by the BFKL equation 
\beq
\mathcal{F}(x,k_\perp, Q_0^2) = \frac{1}{\pi}\delta(k_\perp^2-Q_0^2)
+ \abar \int \frac{d^2q_\perp}{\pi q_\perp^2}&& \!\!\!\!\! \int_x^1 \frac{dz}{z} \biggl ( 
\mathcal{F}(x/z,k_\perp+q_\perp, Q_0^2) - \biggr . \nonumber \\ 
&&\biggl .\theta(k^2_\perp - q^2_\perp) \mathcal{F}(x/z,k_\perp, Q_0^2)
\biggr ),
\label{bfkleq}
\eeq
where $\abar \equiv \alpha_sN_c/\pi$.
This switch implies that the rapidity divergence is cut off since there is an implicit cut in the BFKL formalism. Effectively one introduces a cut-off $\zeta$, and then sets $\zeta=z$ in \eqref{CCHfact}.  

This switch implies a conceptual change which is rather important to understand.
As we have noticed, the would be definition \eqref{LCgluondens} in light-cone gauge corresponds to the WW 
definition. However, once it is instead declared that $\mathcal{F}$ is ``defined'' via the BFKL equation \eqref{bfkleq}, one comes
back to \eqref{bfklfact} and essentially to the dipole definition \eqref{dipgluedistrb2}, and the latter does not 
conform to the parton model idea of a parton distribution.  There is therefore an implicit change in the meaning of $\mathcal{F}$.
Actually, $\mathcal{F}$ is now instead replaced by $F$ in \eqref{bfklfact}, that is the ``BFKL Green's function'' 
which itself is not a ``gluon distribution''.  It is in fact stated in \cite{Catani:1990eg} that for any realistic calculation 
one should rather \emph{provide a non-perturbative $Q_0$-distribution of the 
gluon in the hadron to be convoluted with $\mathcal{F}(x,k_\perp;Q_0^2)$. } What this simply means is that 
one needs to perform a convolution just as in the right hand side of Eq.~\eqref{balitskybfkl} ($Q_0$ here
corresponds to $k_\perp'$ in \eqref{balitskybfkl}) . 
 
It is clear, however, that this cannot be directly compatible with the definition \eqref{LCgluondens} that 
follows from \eqref{CCHfact}, because the operator definition of \eqref{dipgluedistrb2}  cannot
be reduced to \eqref{LCgluondens}  in light-cone gauge.  There is moreover another problem. 
In the set up of \cite{Catani:1990eg} which gives \eqref{CCHfact}, the photon actually scatters off a 
single gluon with momentum $P$. The state $|P\rangle$ in \eqref{dipgluedistrb2}  therefore strictly 
speaking corresponds to an on-shell gluon with momentum $P$, and not to a hadron\footnote{
The cross section in Eq.~\eqref{CCHfact} is therefore really a partonic cross section which is indicated 
by the subscript $\gamma g$.}.  
The idea is 
based on the so-called  ``factorization of mass singularities'' which has  been used 
when dealing with collinear factorization (see for example \cite{pinkbook}).  
In this approach it is first \emph{asserted} 
that a hadronic structure
function, $W^{hadron}$, is  a convolution of the corresponding partonic structure function 
$W^{parton}$ and a so-called ``bare parton density'' $G^{bare}$,
\beq
W^{hadron} (q,P) = W^{parton}(q,p=\xi P) \otimes_\xi G^{bare}(\xi).
\label{massfact1}
\eeq
The convolution here in the longitudinal variable $\xi$ is the same as the $z$ convolution 
in \eqref{CCHfact}. Factorization is then understood as the procedure of extracting out a divergent
factor $D$ from $W^{parton}$ to ``define'' a ``renormalized parton distribution'',
\beq
W^{hadron}  =  (\hat{\sigma} \otimes D) \otimes G^{bare} = \hat{\sigma} \otimes (
D \otimes G^{bare}) = \hat{\sigma} \otimes G^{ren}.
\label{massfact2}
\eeq
For the conceptual problems of this approach in collinear factorization we refer the reader to
\cite{qcdbook}. 
In this case the procedure is thus extended to TMD factorization. 
However, the parameter $Q_0^2$ in  \eqref{bfkleq} implies that  the incoming gluon $P$ is 
no longer on shell, and the type of convolution \eqref{balitskybfkl} which is supposed to 
give a definition of the ``unintegrated gluon distribution'' clearly is rather different than the 
procedure in Eqs.~\eqref{massfact1} and \eqref{massfact2}.  It is also clear that $F$ in \eqref{bfklfact}
is different than $\mathcal{F}$ defined in \eqref{LCgluondens}. 

The question here should be: Can we in the small-$x$ region indeed formulate a formula like 
in \eqref{CCHfact} and then show that it leads to a precise operator definition of the TMD gluon 
distribution which is related to the WW distribution, with all the subtleties regarding the rapidity 
divergences taken properly into account? Additionally, can we then  properly relate this distribution 
to the integrated gluon distribution? Moreover, what evolution equation in rapidity will this distribution 
satisfy? In the TMD approach the rapidity evolution is given by the so-called CSS evolution \cite{qcdbook}, while 
in the formalisms studied here by the BFKL equation. One should investigate the exact connection 
between the two. 

\section{Speakable and unspeakable}

As we have seen above, the idea and concept of the ``unintegrated gluon distribution'' in 
small-$x$ QCD is still intuitively based very much on the parton model definition of a number 
density, but we have also seen that the ``dipole gluon distribution'' does not conform to this idea, 
even if it is intuitively thought of being so (and repeatedly said being so).  The parton model obviously provides a very useful 
intuitive guidance in full QCD, but unnecessary confusion, and even wrong results, can easily arise if one is not careful. 

In the historical development of QCD it was of course the approximative
Bjorken scaling observed in the SLAC-MIT data, together with the successes of the quark-gluon model,
which eventually led to the identification of partons with quarks and gluons, thus providing the latter an 
ontological commitment as real entities (rather than simply being abstract mathematical objects). 

The scaling violations of QCD modify the naive parton model and the definition of the parton distribution
functions which in the parton model are strict number densities according to 
\eqref{eq:pdf.lf.def}.  The parton model is, however, not completely thrown away, but the full QCD result 
rather corresponds to an improvement, without too dramatic or violent differences. Very important in this 
is, however, the existence of factorization theorems in QCD (for a number of elementary processes).  For 
assume that factorization did not hold (its existence requires after all non-trivial proofs). In that case the concept
of parton distributions would not be useful at all, and it would not be possible to gain any real knowledge 
of the underlying entities, the partons, using them. In particular since partons are not directly observable due 
to confinement, we would be in a rather difficult situation in the detailed exploration of the inner structure of hadrons. 
In that case it would obviously be very questionable to commit any ontology to the partons, for example 
speaking of a ``number density'', or a ``$k_\perp$-distribution'' or the ``probability of finding partons with 
momentum fraction between $xP^+$ and $(x+dx)P^+$'' and so on.  Even when factorization does hold, however, 
the strict number density interpretation is lost due to the UV renormalization in QCD, and 
the legitimacy of the statements made upon the parton distributions must be solely based on the 
mathematical rules of QCD. 

Quantum mechanics teaches us to speak of only what we can measure. In accelerator experiments, 
what we measure are a bunch of hadrons and leptons in tracking chambers and calorimeters.  To gain 
any knowledge of the partons from the observed hadrons we employ factorization and parton distributions.
The former is, however, the crucial ingredient in this procedure, which is the moral of the story. 
In short we may say that parton distributions have \emph{no} ontological priority over the factorization 
theorems.  

\section{Summary and Outlook}

The problem of factorization is more intriguing and complex in the case of TMD factorization which 
contains more information on the underlying entities, but which also poses more difficulties. The fact
that factorization appears to be broken in $pp$ collisions  \cite{Collins:2007nk, Rogers:2010dm}
(and mostly likely so in nucleus collisions as well) calls for caution in the applications of naive 
factorization.  As the LHC is a $pp$ (and $AA$) machine at very high energies, it is particularly crucial to understand the
issue in the small-$x$ domain. A preliminary overview of the literature shows
a lack of proofs, and also some confusion as to the meaning of the ``unintegrated gluon distribution''.  

There has recently been many applications of the ``dipole gluon distribution'' in single inclusive hadron 
production (or rather gluon production), see for example \cite{Kharzeev:2004if,  ALbacete:2010ad, Levin:2011hr}. 
The recent multiplicity measurements of ALICE \cite{Aamodt:2010pp} has shown that 
the central multiplicity grows faster with $s$ in $AA$ collisions than in $pp$ collisions.  The question is of course
why this is the case. We must then be very cautious, however, since applications of small-$x$ physics all 
assume the existence of factorization in these processes, and this is highly questionable. Moreover it is 
not at all clear that it is the ``dipole gluon distribution'' which is the relevant object to use in these cases. 
Given the enormous 
complexity especially of the $AA$ collisions, it is no wonder that the naive application of  ``dipole gluon distribution'' 
does not work.  

We have undertaken a comprehensive study \cite{ourpaper} in the hope of provoking further work to address the 
issues mentioned here, so that the important question regarding factorization and the TMD parton distributions
may be hopefully clarified. 

\section*{Acknowledgments}

I would like to thank Anna Stasto and Bowen Xiao for useful discussions.






\bibliographystyle{utcaps}
\bibliography{refs2}

\providecommand{\href}[2]{#2}\begingroup\raggedright\begin{thebibliography}{10}

\bibitem{qcdbook}
J.~C. Collins, ``{Foundations of perturbative QCD, Cambridge University Press
  2011},''.

\bibitem{Collins:2003fm}
J.~C. Collins, ``{What exactly is a parton density?},'' {\em Acta Phys. Polon.}
  {\bf B34} (2003) 3103,
\href{http://arXiv.org/abs/hep-ph/0304122}{{\tt hep-ph/0304122}}.

\bibitem{Johntalk}
J.~C. Collins, ``{Talk given at the QCD evolution workshop: from collinear to
  non collinear case, 8-9 April 2011},''.

\bibitem{Fadin:1975cb}
V.~S. Fadin, E.~A. Kuraev, and L.~N. Lipatov, ``{On the Pomeranchuk Singularity
  in Asymptotically Free Theories},'' {\em Phys. Lett.} {\bf B60} (1975)
50--52.

\bibitem{Kuraev:1977fs}
E.~A. Kuraev, L.~N. Lipatov, and V.~S. Fadin, ``{The Pomeranchuk Singularity in
  Nonabelian Gauge Theories},'' {\em Sov. Phys. JETP} {\bf 45} (1977)
199--204.

\bibitem{Balitsky:1978ic}
I.~I. Balitsky and L.~N. Lipatov, ``{The Pomeranchuk Singularity in Quantum
  Chromodynamics},'' {\em Sov. J. Nucl. Phys.} {\bf 28} (1978)
822--829.

\bibitem{Balitsky:1995ub}
I.~Balitsky, ``{Operator expansion for high-energy scattering},'' {\em Nucl.
  Phys.} {\bf B463} (1996) 99--160,
\href{http://arXiv.org/abs/hep-ph/9509348}{{\tt hep-ph/9509348}}.

\bibitem{Iancu:2002xk}
E.~Iancu, A.~Leonidov, and L.~McLerran, ``{The colour glass condensate: An
  introduction},''
\href{http://arXiv.org/abs/hep-ph/0202270}{{\tt hep-ph/0202270}}.

\bibitem{Dominguez:2011wm}
F.~Dominguez, C.~Marquet, B.-W. Xiao, and F.~Yuan, ``{Universality of
  Unintegrated Gluon Distributions at small x},''
\href{http://arXiv.org/abs/1101.0715}{{\tt 1101.0715}}.

\bibitem{Catani:1990eg}
S.~Catani, M.~Ciafaloni, and F.~Hautmann, ``{High-energy factorization and
  small x heavy flavor production},'' {\em Nucl. Phys.} {\bf B366} (1991)
135--188.

\bibitem{ourpaper}
E.~Avsar and J.~C. Collins, ``{In preparation},''.

\bibitem{pinkbook}
R.~Ellis, J.~Stirling, and B.~Webber, ``{QCD and Collider Physics, Cambridge
  University Press 2003},''.

\bibitem{Collins:2007nk}
J.~Collins and J.-W. Qiu, ``{$k_{T}$ factorization is violated in production of
  high- transverse-momentum particles in hadron-hadron collisions},'' {\em
  Phys. Rev.} {\bf D75} (2007) 114014,
\href{http://arXiv.org/abs/0705.2141}{{\tt 0705.2141}}.

\bibitem{Rogers:2010dm}
T.~C. Rogers and P.~J. Mulders, ``{No Generalized TMD-Factorization in
  Hadro-Production of High Transverse Momentum Hadrons},'' {\em Phys.Rev.} {\bf
  D81} (2010) 094006, \href{http://arXiv.org/abs/1001.2977}{{\tt 1001.2977}}.

\bibitem{Kharzeev:2004if}
D.~Kharzeev, E.~Levin, and M.~Nardi, ``{Color glass condensate at the LHC:
  Hadron multiplicities in pp, pA and AA collisions},'' {\em Nucl.Phys.} {\bf
  A747} (2005) 609--629, \href{http://arXiv.org/abs/hep-ph/0408050}{{\tt
  hep-ph/0408050}}.

\bibitem{ALbacete:2010ad}
J.~L. Albacete and A.~Dumitru, ``{A model for gluon production in heavy-ion
  collisions at the LHC with rcBK unintegrated gluon densities},''
  \href{http://arXiv.org/abs/1011.5161}{{\tt 1011.5161}}.

\bibitem{Levin:2011hr}
E.~Levin and A.~H. Rezaeian, ``{Gluon saturation and energy dependence of
  hadron multiplicity in pp and AA collisions at the LHC},''
  \href{http://arXiv.org/abs/1102.2385}{{\tt 1102.2385}}.

\bibitem{Aamodt:2010pp}
{\bf ALICE} Collaboration, K.~Aamodt {\em et al.}, ``{Charged-particle
  multiplicity measurement in proton-proton collisions at sqrt(s) = 7 TeV with
  ALICE at LHC},'' {\em Eur. Phys. J.} {\bf C68} (2010) 345--354,
\href{http://arXiv.org/abs/1004.3514}{{\tt 1004.3514}}.

\end{thebibliography}\endgroup

\end{document}